\documentclass[%
 reprint,
superscriptaddress,
 amsmath,amssymb,
 aps,
 prl,
]{revtex4-2}

\usepackage{graphicx}
\usepackage{dcolumn}
\usepackage{bm}
\usepackage[colorlinks=true,linkcolor=blue,citecolor=red]{hyperref}

\usepackage{bbm}
\usepackage{physics}
\usepackage{nicefrac}
\usepackage{soul}

\usepackage{color}
  
\usepackage[normalem]{ulem}	

\newcommand{\mrm}{\mathrm}

\newcommand{\sr}[1]{\left[#1\right]}

\newcommand{\kb}{k_{\mathrm{B}}}
\newcommand{\kT}{k_{\mathrm{B}}T}
\DeclareMathOperator{\e}{e}

\newcommand{\dg}{\delta_{\mrm{g}}}
\newcommand{\tildedg}{\tilde{\delta}_{\mrm{g}}}
\newcommand{\ts}{t_{\mrm{s}}}

\newcommand{\fc}{f_{\mrm{c}}}

\newcommand{\fne}{f_\mathrm{ne}}

\newcommand{\Dne}{D_\mathrm{ne}}

\newcommand{\xk}{x_{k}}
\newcommand{\xkp}{x_{k+1}}

\newcommand{\lk}{\lambda_{k}}
\newcommand{\lkp}{\lambda_{k+1}}

\newcommand{\beginsupplement}{%
        \setcounter{table}{0}
        \renewcommand{\thetable}{\arabic{table}}%
        \setcounter{figure}{0}
       \renewcommand{\thefigure}{S\arabic{figure}}
       \setcounter{equation}{0}
       \renewcommand{\theequation}{S\arabic{equation}}
        }

\newcommand{\sfu}{\affiliation{%
 Department of Physics, Simon Fraser University, Burnaby, BC, V5A 1S6 Canada
}}

\newcommand{\hawaii}{\affiliation{%
 Department of Physics and Astronomy, University of Hawaii at M\=anoa, Honolulu, HI 96822, USA
}}

\begin{document}

\preprint{APS/123-QED}

\title{Information engine in a nonequilibrium bath}

\author{Tushar K.\ Saha}
\thanks{These authors contributed equally.}
\sfu
\author{Jannik Ehrich}
\thanks{These authors contributed equally.}
\sfu
\hawaii
\author{Mom\v{c}ilo Gavrilov}
\sfu
\author{Susanne Still}
\hawaii
\author{David A.\ Sivak} 
\sfu
\author{John Bechhoefer}
\sfu

\date{\today}

\begin{abstract}
Information engines can convert thermal fluctuations of a bath at temperature $T$ into work at rates of order $\kT$ per relaxation time of the system. We show experimentally that such engines, when in contact with a bath that is out of equilibrium, can extract much more work. We place a heavy, micron-scale bead in a harmonic potential that ratchets up to capture favorable fluctuations. Adding a fluctuating electric field increases work extraction up to ten times, limited only by the strength of applied field. Our results connect Maxwell’s demon with energy harvesting and an estimate of efficiency shows that information engines in nonequilibrium baths can greatly outperform conventional engines. 

\end{abstract}

\maketitle

Maxwell's famous thought experiment proposed a way to convert information about thermal fluctuations into energy~\cite{maxwell1872}. Exploring ``Maxwell demons'' has improved our understanding of the second law of thermodynamics~\cite{leff2002maxwell,parrondo2015}. Building them, as ``information engines'', a concept inspired originally by Szilard's model~\cite{Szilard1929}, has allowed tests of the second law applied to mesoscopic length scales~\cite{toyabe2010, koski2014pnas, chida2017power, paneru2018, ribezzi2019, cottet2017}. The ability of an information engine to extract work from a single heat bath is reconciled with the second law because the cost of sensing fluctuations and exploiting the relevant information equals or exceeds the energy extracted~\cite{Szilard1929, bennett1982, parrondo2015, Horowitz2014a}.

That information-processing costs compensate for the extracted energy implicitly assumes that the measuring device operates at the same temperature as the engine itself.  If the temperature of the engine bath exceeds that of the bath connected to the measuring device, net work can be extracted~\cite{still20}. But the range of temperatures available for heat baths is small, which limits engine power, even with optimized information processing~\cite{StillDaimer2022}. For example, the ratio of boiling-water to freezing-water temperatures is $373/273 \approx 1.4$, and the ratio of temperatures in an internal-combustion heat engine is $\lesssim 8$, with practical efficiencies $\lesssim 0.5$~\cite{ghojel20}.

In this paper, we show experimentally that this limitation can be overcome by immersing the information engine in a nonequilibrium heat bath. The environment is out of equilibrium at macroscopic scales but has practically unchanged local temperature $T$. The measuring device of the engine is in contact with an equilibrium heat bath also at temperature $T$. Such an information engine extracts energy from both thermal and nonequilibrium fluctuations. Similar ideas have been proposed theoretically to take advantage of active fluctuations produced by bacteria swimming in a bath~\cite{paneru22} and to increase the output of a Szilard engine~\cite{malgaretti22}. 

With carefully chosen experimental parameters, our engine can extract work at up to ten times the maximum rate achievable when connected to an equilibrium bath at temperature $T$. Moreover, we show that energy extraction is constrained only by practical experimental limits on the nonequilibrium forcing of the external environment. The energy extracted can, in principle, exceed the costs associated with necessary information processing and control by orders of magnitude.

\begin{figure}[ht]
    \centering
    \includegraphics[width=0.8\linewidth]{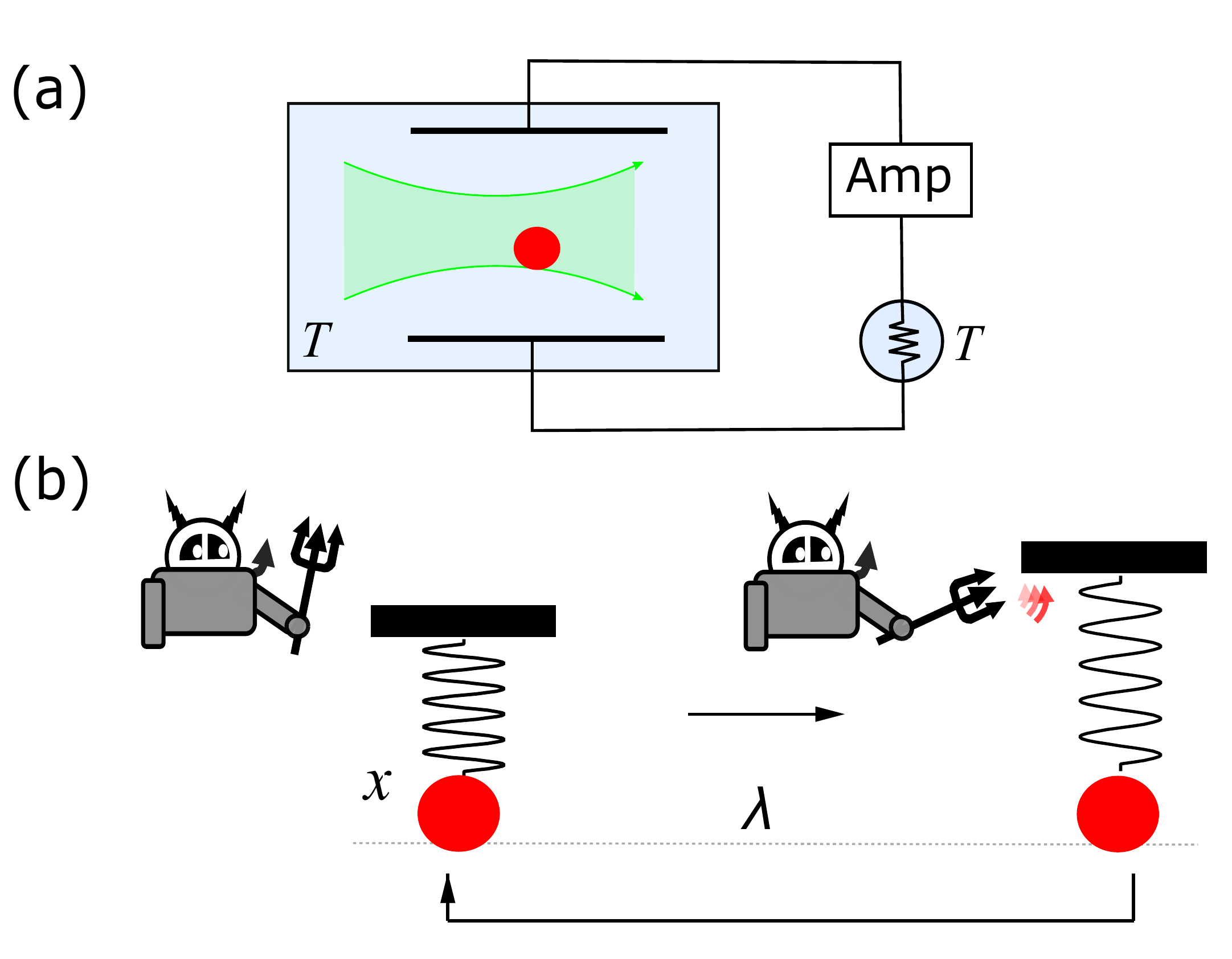}
    \caption{ Schematic of information engine in a nonequilibrium bath. (a) Optically trapped bead in water, subject to nonequilibrium external noise. This external noise is generated by electrodes connected to a resistor in a thermal bath, followed by an amplifier. Both bead and resistor are at room temperature $T$. (b) Schematic of the information engine.}
    \label{fig:engine_schematic}
\end{figure}

\textit{Experimental setup.}---The information engine consists of an optically trapped, heavy bead in water that acts as a thermal bath at room temperature $T$~\cite{saha2021maximizing}. To force the bath out of equilibrium, a fluctuating electric field is applied via electrodes~\cite{martinez13}, as shown in Fig.~\ref{fig:engine_schematic}(a).  Unlike previous experimental implementations of baths with higher effective temperatures that were based on digitally generated noise~\cite{martinez13,chupeau18,militaru21,goerlich21}, the fluctuating field here arises from a physical reservoir, the amplified Johnson voltage noise of a resistor at temperature $T$. The nonequilibrium-noise strength is regulated by a variable-gain amplifier, whose bandwidth is set by a low-pass filter with adjustable cutoff frequency.  Under gravity, the bead fluctuates about a mean position because of thermal and nonequilibrium forces acting on it. Further details on the setup are given in the Appendix.

\textit{System dynamics.}---We model the trapped bead as a spring-mass system, as shown in Fig.~\ref{fig:engine_schematic}(b). The information engine operates by raising the trap position when the bead fluctuates above the trap center. The upward fluctuation of the bead, increasing its gravitational potential energy, is rectified by ratcheting the trap position. We update the trap position to convert the thermal fluctuations from the nonequilibrium bath into gravitational energy, carefully choosing the distance the trap is moved, so that the shifted trap potential does not perform work on the bead~\cite{saha2021maximizing}. 

The position $x(t)$ of a bead trapped in a harmonic potential centered on $\lambda$ at time $t$ in a nonequilibrium bath is described by the Langevin equation
\begin{align}
    \dot{x}(t) &=\underbrace{- \left[x(t)-\lambda(t)\right]}_{\text{restoring force}} \, - \underbrace{\dg}_{\text{grav.}} +\underbrace{\xi(t)}_{\text{thermal}}+\underbrace{\zeta(t)}_{\text{noneq.}} , 
\label{eq:sto_EQM_full}
\end{align}
where we have rescaled lengths by the bead position's equilibrium standard deviation $\sigma \equiv \kb T/\kappa$ in the trap with strength $\kappa$, and times by the relaxation time $\tau_\mathrm{r} \equiv \gamma / \kappa$ of a bead with Stokes' friction coefficient $\gamma$ in the trap. The scaled effective mass $\dg \equiv \Delta m g/(\kappa\sigma)$ accounts for the effects of gravity and buoyancy on a bead with effective mass $\Delta m = \Delta \rho (4/3)\pi r^3$, with $r$ the particle radius and $\Delta \rho$ the density difference between particle and the surrounding fluid. The noise $\xi(t)$ reflects equilibrium thermal fluctuations of the water bath and is modeled by Gaussian white noise with zero mean and variance $\ev{\xi(t)\xi(t')} = 2\, \delta(t-t')$.

We measure the bead position at a sampling time $\ts = 20$ \textmu s, using the forward-scattered light from a detection laser. The trap is updated at the same time that measurements are made, responding to measurements that took place 20 \textmu s in the past. The ratchet feedback algorithm is
\begin{align}
    \lkp = \lk+\alpha\,(\xk-\lk)\,\Theta(\xk-\lk)\,,
\label{eq:trap-center}
\end{align}
for Heaviside (step) function $\Theta(\cdot)$ and scalar feedback gain $\alpha$.

\textit{Nonequilibrium bath.}---The second noise $\zeta(t)$ in Eq.~\eqref{eq:sto_EQM_full} describes random electrokinetic forcing of strength $\Dne$. The electrokinetic forces combine electroosmotic and electrophoretic effects on the bead~\cite{cohen06}. Here, we empirically determine the nonequilibrium-noise strength.

The fluctuating electrokinetic forces arise from the amplified, low-pass-filtered Johnson noise of a resistor. In scaled units, the noise term obeys
\begin{align}
    \fne^{-1}\,\dot{\zeta}(t) = -\zeta(t) + \sqrt{\Dne}\, \tilde{\xi}(t)\,,
\label{eq:OU_noise_LE}
\end{align}
where $\fne$ is the \textit{cutoff} frequency set by the filter frequency, $\ev{\tilde{\xi}(t)}=0$, and $\ev{\tilde{\xi}(t) \tilde{\xi}(t')} = 2\,\delta(t-t')$. The low-pass filter generates exponentially correlated colored nonequilibrium Ornstein-Uhlenbeck noise, $\ev{\zeta(t)\zeta(t')} = \Dne\,\fne \e^{-\fne |t-t'|}$, which tends to white noise for $\fne \to \infty$. The dynamics of particles subjected to such noise have been studied extensively~\cite{Haenggi1995_Colored,Szamel2014_Self-propelled,Goerlich2021_Harvesting,Ghosh2022_Statistical}. In experiments, we can vary $\Dne$ from 0 to 83.5 and $\fne$ from 10~Hz to 24~kHz. The trap's cutoff frequency $\fc = 1/(2\pi \tau_\mathrm{r})$ is 200 $\pm$ 2~Hz.

\textit{Energy measurements.}---
At each time step, the stored gravitational (free) energy of the bead changes by
\begin{align}
\label{eq:freeEnergy}
    \Delta F_{k+1} = \dg\,\left(\lkp-\lk\right) \,.
\end{align}
The average output power is measured using $\dot F~ =~\sum_{k=0}^N \ev{\Delta F_k} /N\ts$, where the ensemble average $\ev{\cdot}$ is estimated by averaging over multiple trajectories, each of length $N$ time steps.

The work done by the trap on the bead is
\begin{align}
    W_{k+1} = \tfrac{1}{2} \sr{(\xkp-\lkp)^2 - (\xkp-\lk)^2}\,,
\label{eq:trapWork}
\end{align}
and the average trap power is measured using $\dot W~=~\sum_{k=0}^N \ev{W_k} /N\ts$.

Equation~\eqref{eq:trapWork} implies that the trap power would be zero for $\alpha = 2$ if there were no feedback delay and no measurement uncertainty~\cite{saha2021maximizing}; however, the one-step delay implies that the zero-work condition is realized at a lower feedback gain, empirically $\alpha \approx 1.8$, when measurement noise is negligible~\cite{saha2022bayesian}.

\textit{Results.}---We study the dependence of the information engine's output power on the characteristics of the nonequilibrium noise. 

\begin{figure}[ht] 
    \centering
    \includegraphics[width=1\linewidth]{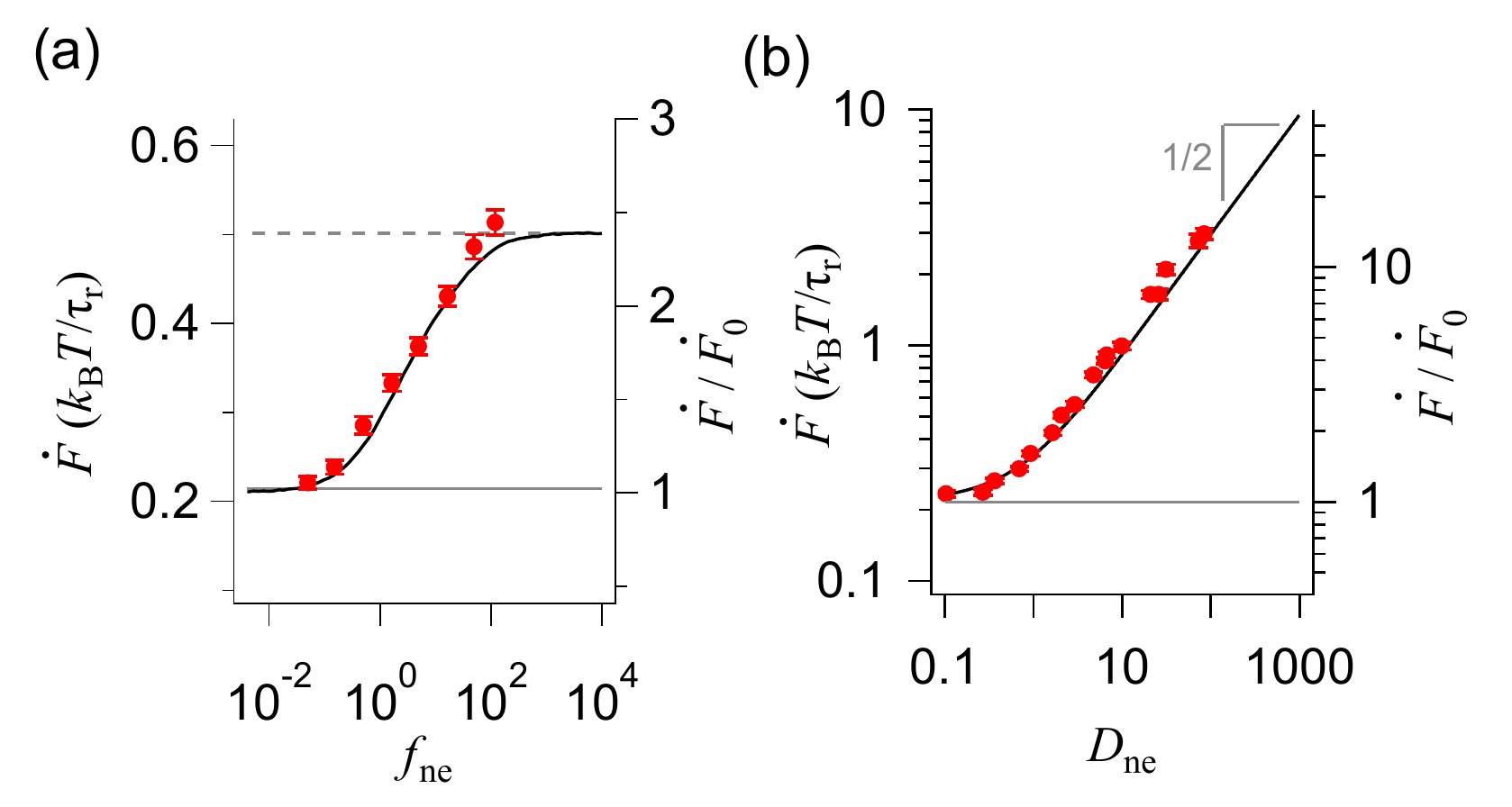} 
    \caption{Output-power optimization for a 3-\textmu m bead at $\alpha = 1.8$. (a) Output power as a function of cutoff frequency $\fne$ of the nonequilibrium bath scaled by the trap cutoff frequency $\fc$, at nonequilibrium-noise strength $\Dne = 3.0$ and scaled mass $\dg=0.37$. Dashed line: high-$\fne$ limit, from Eq.~\eqref{eq:power_white_noise}. Gray line: measured output power $\dot{F}_0$ at equilibrium ($\Dne~=~0$). Black curve: prediction based on numerical simulations, using the measured $\Dne$; not a fit. Right axis: ratio of nonequilibrium to equilibrium output powers. (b) Output power as a function of $\Dne$ for high cutoff frequency ($\fne = 118$). Gray line: measured output power at $\Dne = 0$ and $\dg = 0.38$. Black curve: scaling-theory prediction from Eq.~\eqref{eq:power_white_noise}, showing the asymptotic $\sim \Dne^{1/2}$ dependence.}
\label{fig:Fvsfc}
\end{figure}

First, we measure the output power $\dot F$ for different noise cutoff frequencies $\fne$, at fixed amplitude $\Dne = 3.0$.
Figure~\ref{fig:Fvsfc}(a) shows that the output power increases with $\fne$, in agreement with numerical simulations~(see Appendix). The output power saturates at $\fne \gtrsim 100$:  At high $\fne$, the bead cannot follow the force fluctuations and thus effectively experiences white noise indistinguishable from that of a thermal bath with a higher ``effective temperature.'' Further increases in the cutoff frequency do not affect the bead's dynamics. At low $\fne$, the nonequilibrium fluctuations are weaker than the equilibrium thermal fluctuations, and the output power equals that in a thermal bath at room temperature. At the maximum $\fne$, the output power for $\Dne=3.0$ is more than twice (2.4$\times$) that for a purely thermal bath ($\Dne=0$).

Next, we study the dependence of the output power on the nonequilibrium-noise strength $\Dne$ using a variable-gain amplifier. We fix the low-pass filter's cutoff frequency $\fne$ to be $\approx 100$ times
that of the trap, so that all experiments are in the limit where the nonequilibrium environment provides effectively white-noise fluctuations to the bead. Figure~\ref{fig:Fvsfc}(b) shows that the output power increases with the noise strength $\Dne$. At low $\Dne$, the power is that achievable with a purely thermal bath at room temperature. As $\Dne$ is increased, the output power increases monotonically. In our experiments, electrochemical reactions at the electrodes limit the maximum achievable noise strength and hence the output power of the information engine. Maximizing both cutoff frequency $\fne$ and noise strength $\Dne$, we achieve an increase of 14$\times$ in output power relative to the equilibrium case, to $3.8\times10^{3}\, k_\textrm{B}T$/s = $1.6\times10^{-17}$ W.

A simple scaling argument in the white-noise limit explains the observed performance increase: In Ref.~\cite{saha2021maximizing}, we found the output power of the purely thermal information engine to be, for $\ts \ll 1$ and with $\dot{F}_0 \equiv \dot F(\Dne = 0)$,
\begin{align}
    \dot F_0 &= \sqrt{\frac{2}{\pi}}\dg \e^{-\dg^2/2} \left[1 + \erf\left( \frac{\dg}{\sqrt{2}} \right) \right]^{-1} \,.
\label{eq:power_thermal}
\end{align}

For $\fne \gg 1$, the exponentially correlated Ornstein-Uhlenbeck noise $\zeta(t)$ in Eq.~\eqref{eq:OU_noise_LE} becomes effectively white noise, relative to the trap's cutoff frequency $\fc$. In this limit and ignoring inertial and hydrodynamic corrections to the overdamped Langevin dynamics of the bead~\cite{franosch11}, it makes sense to view the bead as being immersed in a bath with a higher ``effective temperature'' $T_\mathrm{ne}$. Since $\xi$ and $\tilde{\xi}$ are uncorrelated,
\begin{multline}
    \left\langle \left[ \xi(t) + \sqrt{\Dne} \, \tilde{\xi}(t)\right] \left[\xi(t') + \sqrt{\Dne} \tilde{\xi}(t')\right] \right\rangle\\
    = 2 (1+\Dne)\delta(t-t')\,,
\end{multline}
and the effective temperature ratio, in physical units, is $T_\mathrm{ne}/T=1+\Dne /D$ or, in scaled units, $T_\mathrm{ne}=1+\Dne$.

The higher effective temperature $T_\mathrm{ne}$ affects the scaling of output power and, via the length scaling $\sigma(T) = \sqrt{\kb T/\kappa}$, the scaled effective mass,
\begin{align} 
    \tildedg(\Dne) = \dg\, \frac{\sigma(1)}{\sigma(1+\Dne)} = \frac{\dg}{ \sqrt{1+\Dne}}\,.
\label{eq:dg_noneq_noise}
\end{align}
Therefore, in the white-noise limit, the output power is
\begin{align}
    \dot F = (1\!+ \!\Dne) \sqrt{\frac{2}{\pi}} \, \tildedg \, \e^{-\tildedg^2/2} \left[ 1 + \erf \left( \frac{\tildedg}{\sqrt{2}} \right) \right]^{-1} \,.
\label{eq:power_white_noise}
\end{align}
Substituting Eq.~\eqref{eq:dg_noneq_noise} for $\tildedg$ into Eq.~\eqref{eq:power_white_noise} gives, for $\Dne~\gg~1$, the asymptotic behavior $\dot F \sim \Dne^{\,1/2}$ seen in Fig.~\ref{fig:Fvsfc}(b).

\begin{figure}[htb]
    \centering
    \includegraphics[width=0.7\linewidth]{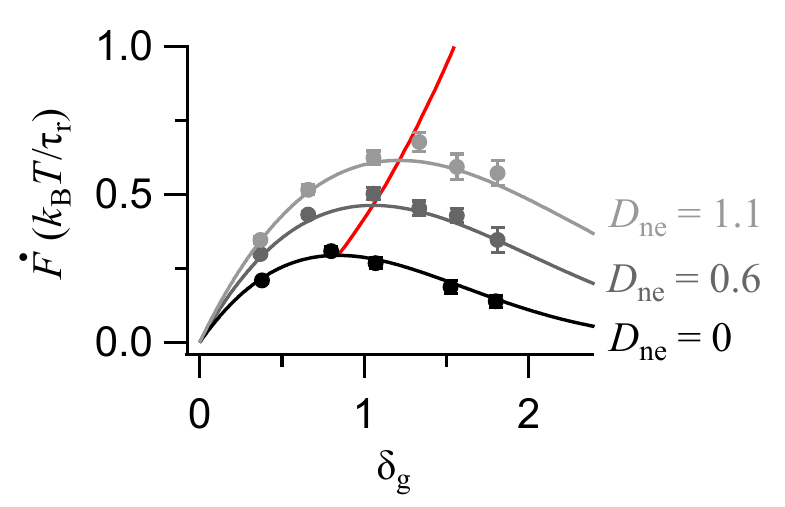}
    \caption{Optimizing output power as a function of scaled mass $\dg$ for different nonequilibrium-noise strengths. Markers: experiments; lines: Eq.~\eqref{eq:power_white_noise}. Red curve parametrically plots maximal $\dot{F}$ and $\dg$ that maximizes $\dot{F}$ (calculated from Eq.~\ref{eq:dgmax}), as functions of $\Dne$.}
    \label{fig:Fvsdg}
\end{figure}

Finally, we study the dependence of the output power on the scaled mass $\dg$. Figure~\ref{fig:Fvsdg} shows that for $\Dne = 0$, the output power is maximized at an optimum scaled mass $\dg \approx 0.845$~\cite{saha2021maximizing}. The optimum arises from a trade-off: having a larger mass increases the gravitational energy gained from a favorable up-fluctuation but reduces the frequency of such fluctuations~\cite{saha2021maximizing,Lucero2021_Maximal}.

Figure~\ref{fig:Fvsdg} shows that the optimal mass $\dg$ increases with the nonequilibrium-noise strength $\Dne$. The nonequilibrium noise increases the amplitude of the bead's fluctuation and makes ratchet events more frequent, thereby shifting the optimal trade-off to higher $\dg$. Comparing Eqs.~\eqref{eq:power_thermal}~and~\eqref{eq:power_white_noise}, the maximum output power is achieved for $\tildedg(\Dne) \approx 0.845$, and hence, according to Eq.~\eqref{eq:dg_noneq_noise}, for 
\begin{align}
    \dg \approx 0.845\sqrt{1+\Dne} \,.
\label{eq:dgmax}
\end{align} 
Equation~\eqref{eq:dgmax}, the red curve in Fig.~\ref{fig:Fvsdg}, represents the maximum achievable output power for different nonequilibrium-noise strengths $\Dne$, achieved at optimal $\dg$.

\textit{Efficiency.}---As we have seen, operating an information engine in a nonequilibrium bath with noise strength $\Dne$ (and effective temperature $T_\mathrm{ne}$ in the white-noise limit) can increase output power. At the same time, the measuring device and controller that gather and exploit the information used to power the engine are in contact with an equilibrium bath at temperature $T$ and thus independent of the nonequilibrium driving force $\Dne$. 

We therefore expect the ratio $\Upsilon$ of these two powers (power extracted over minimum operating power), one possible measure of ``efficiency,'' to increase with $\Dne$. (Note that $\Upsilon$ differs from the Carnot efficiency and, as we will see, is not bounded by unity.) To estimate this increase, we consider the minimum additional work needed to run the controller, which equals the reduction due to the controller's dynamics in the conditional entropy~\cite{Cover2006_Elements} $H[\Lambda|X]$ of the trap position $\lambda$ given the particle position $x$~\cite{Ehrich2022_Energetic}. Therefore the \emph{information power} required to measure, erase information, and control the engine is
\begin{align}
     P_\mathrm{info} \equiv \frac{H[\Lambda_{k-1}|X_k] - H[\Lambda_k|X_k]}{\ts}\,.
\label{eq:Pinfo}
\end{align}
In the white-noise limit ($\fne\to\infty$), the two conditional entropies can be estimated from simulations and analytical approximations~(see Appendix). Figure~\ref{fig:efficiency}(a) compares the input information power $P_\mathrm{info}$ with the output rate $\dot F$ of free-energy gain, Eq.~\eqref{eq:power_white_noise}, as a function of the noise strength $\Dne$. While the engine output grows as $\Dne^{1/2}$, the input power saturates for large $\Dne$, in principle permitting extraction of orders of magnitude more power than required to run the engine, because larger noise strength offers more fluctuations to rectify but does not affect the controller.

Figure~\ref{fig:efficiency}(b) shows the engine's efficiency $\Upsilon \equiv \dot{F}/P_\mathrm{info}$, the ratio of output to input powers.  With sufficiently strong nonequilibrium fluctuations, output power can exceed input power. In contrast, a ``conventional'' engine that drags the particle upwards against gravity at the same velocity $v=\dot F/\dg$ requires trap power $\dot W = v^2 + \dg v$ and has efficiency $<1$~(see Appendix). This conventional strategy is more efficient at low output power, illustrating that rectifying purely thermal fluctuations is inefficient when using a measuring device and controller that operate at the same temperature. However, with sufficiently strong nonequilibrium fluctuations the information ratchet can extract energy more efficiently. Experimentally, we do not reach the regime of larger output than input power; however, with the largest experimental noise strengths, the information ratchet is more efficient than the conventional dragging strategy.  Increasing the electric field strengths by using more closely spaced electrodes and more careful choices of electrode material and bath composition could substantially increase the achievable $\Dne$ and lead to output powers that exceed minimum information-processing costs.

\begin{figure}[tb]
    \centering
    \includegraphics[width=1\linewidth]{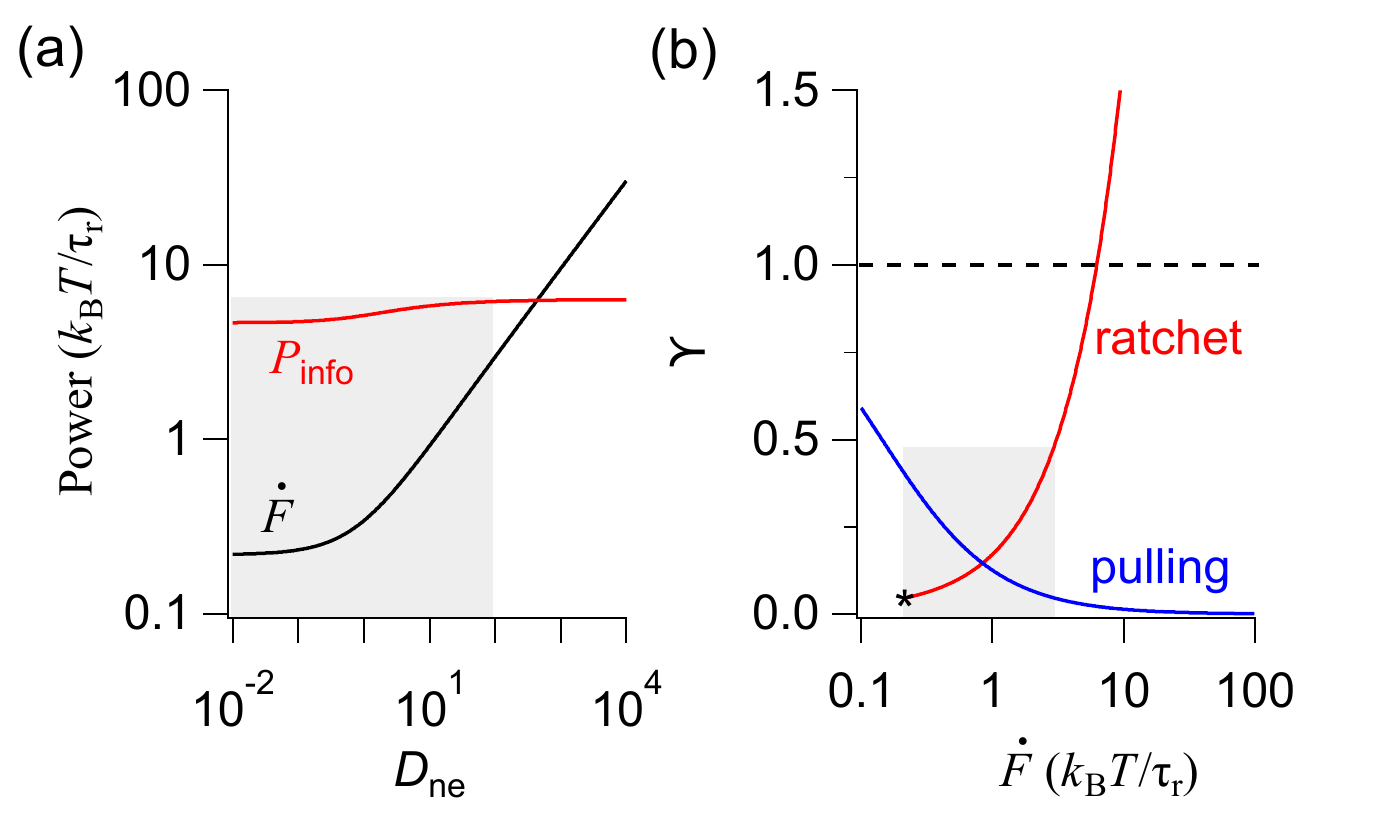}
    \caption{Numerical estimates of information-engine efficiency in the white-noise limit ($\fne=10^4$). (a) Output free-energy gain $\dot F$ [Eq.~\eqref{eq:power_white_noise}] and information power $P_\mathrm{info}$ (minimum power to perform measurement, erase information, and control the information engine) as a function of nonequilibrium-noise strength $D_{\rm ne}$. (b) Efficiency $\Upsilon$ as a function of $\dot F$, compared with a pulling experiment that drags the particle upwards at constant velocity, instead of ratcheting. Star indicates the equilibrium bath, $\Dne=0$. In all calculations, $\dg=0.38$ and $\ts=1/40$ (experimental parameters). Shaded regions indicate experimentally accessible nonequilibrium-noise strengths.} 
    \label{fig:efficiency}
\end{figure}

\textit{Discussion and Conclusion.}---We have shown experimentally that an information engine in contact with a nonequilibrium bath can extract and store an order of magnitude greater power than it can in the same bath without active fluctuations. If the measuring and control devices are at the nominal temperature of the bath (without external forcing) and if the forcing of the nonequilibrium bath is sufficiently strong, then more energy can be extracted than the minimum energy needed to run the measurement-and-control system. Our experiments achieved an efficiency of
48.5\%, limited only by the amount of forcing we were able to supply the bath.

To understand the significance of these experimental results, we recall that information engines can be connected to two heat baths, a higher-temperature one for the engine itself and a lower-temperature one for the measuring device and controller~\cite{still20}. This generalized information engine framework was used recently to automate not only the demon's function \cite{Szilard1929} but also its information processing~\cite{still20}. Such engines can produce net positive work output and, with optimal information processing, have efficiencies that approach the Carnot limit~\cite{StillDaimer2022}. Similarly, passive ratchets can extract net work only when connected to heat baths at different temperatures~\cite{Smoluchowski1912,feynman63,parrondo96}.

As discussed in the Introduction, typical temperature ratios are $\lesssim 10$; however, although an information engine is limited by the Carnot efficiency associated with the ratio of (effective) temperatures of the two baths, it extracts energy from only a very small fraction of modes of the high-temperature bath. In particular, as Fig.~\ref{fig:Fvsfc}(a) shows, an engine need only be supplied with modes slightly exceeding $\fc$ (e.g., $\fne \sim 10^3$ Hz) to achieve half the maximum possible output. By contrast, equipartition implies that a bath in equilibrium at a temperature $T$ has modes with equal energies up to phonon frequencies, $k_\textrm{B}T/h \approx \mathcal{O}(10^{13})$~Hz, with $h$ Planck's constant~\cite{nyquist28}. Thus, the fraction of forced modes is only $10^{-10}$. Put another way, if the nonequilibrium forcing is removed and the bath returns to equilibrium, its temperature does not measurably increase.

Engines can thus extract work from nonequilibrium modes, while the associated measuring device remains at equilibrium. Given a white-noise spectrum with frequencies 10--100$\times$ higher than the engine cutoff frequency, we can assign an effective temperature to the nonequilibrium bath. Because work extraction draws on so few modes, the effective temperature of the bath can be orders of magnitude higher than physical temperatures.

Very high effective temperature and correspondingly large work-extraction rates are implicit in many old technologies: Sailboats move because their sails are constantly adjusted to catch the wind, and wind turbines similarly generate power by adjusting their rotors to be normal to the fluctuating wind direction~\cite{stavrakakis12}. On a smaller scale, self-winding watches, first developed in the 18th century~\cite{watkins16}, rectify the nonequilibrium fluctuations supplied by  movements of the wearer's arm~\cite{parrondo96}. More recent experimental realizations include ratchets driven by granular gasses~\cite{Eshuis2010_Experimental, Joubaud2012_Fluctuation, Gnoli2013_Brownian, Gnoli2013_Nonequilibrium, lagoin22} that achieve  effective temperatures up to $10^{15} \times$ room temperature~\cite{Rouyer2000_Velocity,Feitosa2004_Fluidized,Chastaing2017_Two_Methods} and rotors driven by turbulence~\cite{Francois2020_Nonequilibrium}.

In the above examples, the scale of the system correlates with the power extracted. Our experiments use micron-scale beads and extract powers of 2.97~$k_\mathrm{B}T/\tau_\mathrm{r} \approx 10^{-17}$~W.  For granular media, millimeter-sized beads lead to extracted powers of $10^{-6}$~W~\cite{lagoin22}. For wind turbines, the 100-m scale blades lead to extracted powers of $10^6$~W~\cite{mathew12}. Thus, larger length scales increase the power that can be extracted from a fluctuating environment.

Nonequilibrium fluctuations can also be generated by active media~\cite{Ramaswamy2010_Mechanics, Marchetti2013_Hydrodynamics,Sokolov2010_Swimming, Leonardo2010_Bacterial, Reichhardt2017_Ratchet, Vizsnyiczai2017_Light,Pietzonka2019_Autonomous,Fodor2021_Active,Speck2016_Stochastic,Fodor2016_How,Mandal2017_Entropy,Pietzonka2018_Entropy,Dabelow2019_Irreversibility,Caprini2019_Entropy,Dabelow2021_Irreversibility} such as (suspensions of) microswimmers~\cite{Elgeti2015_Physics} and active Brownian particles~\cite{Bechinger2016_Active}. The vastly higher values of effective temperature that one can achieve relative to physical temperatures suggest the potential for drastic efficiency increases. Indeed, we have seen that the work extracted can in principle exceed the minimum information costs associated with the engine function. We then speculate that exploitation of fluctuations, as shown here, could be an organizing principle for molecular machinery, where strong nonequilibrium fluctuations~\cite{Mizuno2007_Nonequilibrium,Gallet2009_Power} have been shown to speed up various cellular processes~\cite{Ariga2021_Noise-Induced, Tripathi2022_Acceleration}.

Finally, our results also highlight a different way to understand energy harvesting~\cite{Beeby2006_Energy,Priya2007_Advances,Cook-Chennault2008_Powering,Mitcheson2008_Energy} by microscopic devices. Such analyses are often specific to the type of systems analyzed. For example, mechanical energy harvesters depend heavily on resonant-forcing mechanisms that make inefficient use of the spectrum of fluctuations~\cite{Gammaitoni2011_Vibration}. Our approach gives maximum estimates (for a given fluctuation spectrum) of the power that can in principle be extracted and can thus serve as benchmarks for existing systems and may suggest new extraction strategies. The question, which has only begun to be addressed in special cases, is whether ``intelligently chosen interventions''~\cite{Liu2019_Intelligently} can outperform standard passive rectification strategies, such as full-wave rectifier bridge circuits based on diodes or Brownian ratchets~\cite{Lopez2008_Realization}.

\begin{acknowledgements}
We thank Joseph Lucero for helpful conversations. This research was supported by grant FQXi-IAF19-02 from the Foundational Questions Institute Fund, a donor-advised fund of the Silicon Valley Community Foundation. Additional support was from grant FQXi-RFP-1820, co-sponsored with the Fetzer Franklin Fund (S.S.), Natural Sciences and Engineering Research Council of Canada (NSERC) Discovery Grants (D.A.S.\ and J.B.) and a Tier-II Canada Research Chair (D.A.S.).
\end{acknowledgements}

\section*{Appendix}
\appendix
\beginsupplement

\section{Experimental Apparatus}

\begin{figure}[ht!]
    \centering
    \includegraphics[width=1\linewidth]{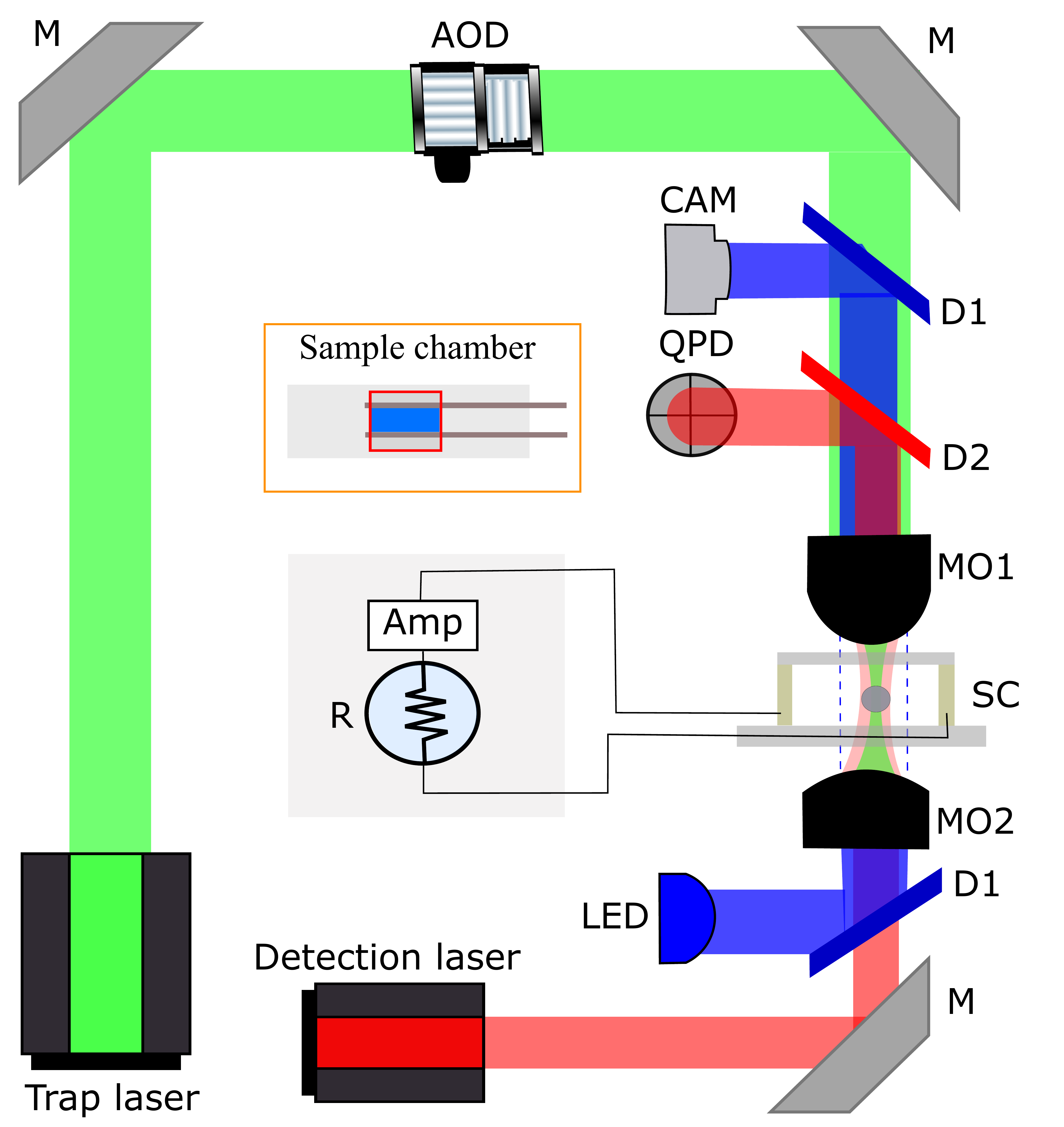}
    \caption{Schematic diagram of the experimental apparatus with external electrical noise (shaded gray region). R = resistor, Amp = amplifier, M = mirror, AOD = acousto-optic deflector, D1 = blue dichroic filter, D2~=~red dichroic filter, SC = sample chamber, MO1 = trapping microscope objective, and MO2~=~detection microscope objective. Inset (orange box): sample chamber with two copper-tape electrodes (brown lines). The blue shaded region represents the bead solution.}
    \label{fig:apparatus}
\end{figure}

Figure~\ref{fig:apparatus} shows the schematic diagram of the apparatus. A bead is trapped by focusing a horizontally aligned (perpendicular to gravity) green trapping laser (H\"UBNER Photonics, Cobolt Samba, 1.5 W, 532 nm), using a microscope objective (MO1, water-immersion 60x Olympus objective 1.2 NA).

A red laser (H\"UBNER Photonics, Cobolt Flamenco, 100 mW, 660-nm) is used for bead-position detection. The scattered detection light from the bead is collected using another microscope objective (MO2, 40x Nikon, 0.5 NA) and focused on the quadrant photo-diode (QPD, First Sensor, QP50-6-18u-SD2) to measure the bead position. The position of the trap is changed using acousto-optic deflectors (AODs, DTSXY-250-532, AA Opto Electronic), one for each axis perpendicular to the direction of laser propagation. We use the AOD that controls the axis parallel to gravity, pointing out-of-plane, to perform the experiments. References~\cite{saha2021maximizing, kumar2018nanoscale} provide more details about the apparatus and the calibration.

\section{Nonequilibrium noise source}

The nonequilibrium energy is generated by connecting the sample chamber to an external electronic-noise source. The sample chamber was built using a glass slide and a coverslip, which were separated by double-layered sticky copper tape. Long strips of copper tape extend beyond the glass slide and connect to external electrodes using alligator clips (see Fig.~\ref{fig:apparatus}, orange inset). The space between the electrodes is $\approx 5$ mm. 

\begin{figure}[ht!]
    \centering
    \includegraphics[width=0.8\linewidth]{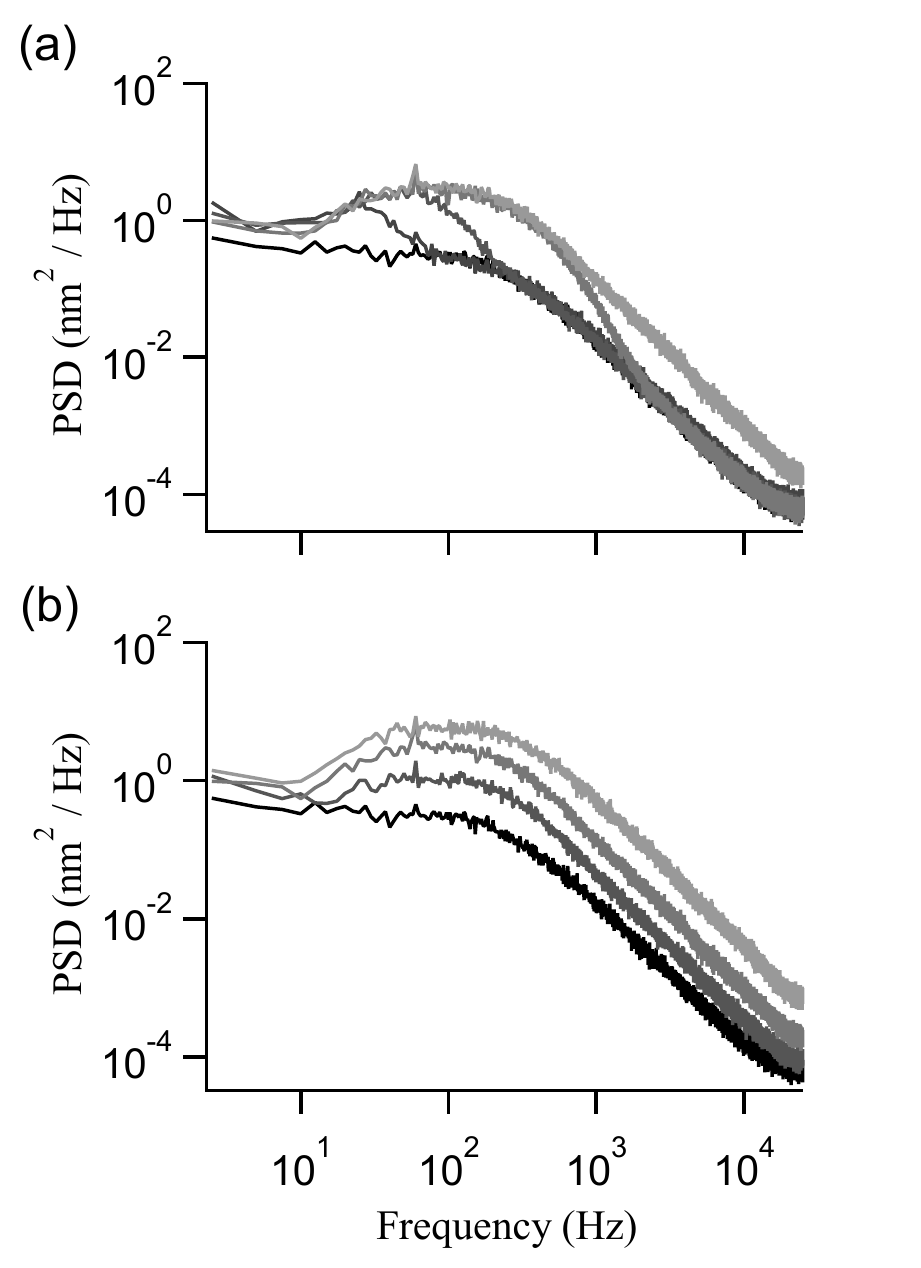}
    \caption{Power spectral density of a trapped bead in the presence of external noise. Black curve: power spectrum in the absence of nonequilibrium noise. (a) Lighter shades of gray: fixed noise amplitude (variable-amplifier gain $G_2 = 6\times10^2$) at higher cutoff frequencies of the noise: 30, 100, 1000, and 100\,000 Hz. (b) Lighter shades of gray: fixed cutoff frequency (100\,000 Hz) with increasing variable-amplifier gain, with $G_2$~=~\{3, 6, 10\}~$\times10^2$.}
    \label{fig:PSD}
\end{figure}

To generate the stochastic forces, we use the Johnson noise of a 300-k$\Omega$ resistor (TeachSpin, Noise Fundamentals). The TeachSpin noise source has a built-in, two-stage preamplifier whose overall gain is $G_1 = 600$. The noise is further amplified using two other external amplifiers: First, there is a variable-gain amplifier ($G_2=10$--$10^4$) using a TeachSpin noise controller. Second, there is a home-built, fixed-gain ($G_3 = 15$) amplifier that can source the higher current that is needed for this experiment (the two copper electrodes form a capacitor that charges more quickly with higher currents). 

We control the bandwidth of the noise using high-pass and low-pass two-pole Butterworth filters included with the TeachSpin controller. The high-pass filter cutoff frequency is set to 10~Hz to prevent electroconvection in the fluid and low-frequency electronic noise. Its bandwidth is sufficiently small compared to the cutoff frequency of the nonequilibrium noise (24~kHz) that calculations of fluctuation strengths that extend to zero frequency do not differ significantly from the experimental results. Experiments at the lowest three cutoff frequencies (10, 30 and 100 Hz) in Fig. 2 (a) were performed using the built-in high-pass filter of the TeachSpin amplifier with cutoff frequency of 2.5~Hz. 

The power spectra of trapped beads at different low-pass frequency cutoff and amplifier gain are presented in Fig.~\ref{fig:PSD} (a) and (b), respectively. Experiments were performed for parameter values that did not generate electrochemical breakdown of the electrodes. The copper-tape electrodes were used for simplicity; replacing them with platinum electrodes would increase the range of voltages that can be applied to the bead. 

\begin{figure}[ht!]
    \centering 
    \includegraphics[width=0.8\linewidth]{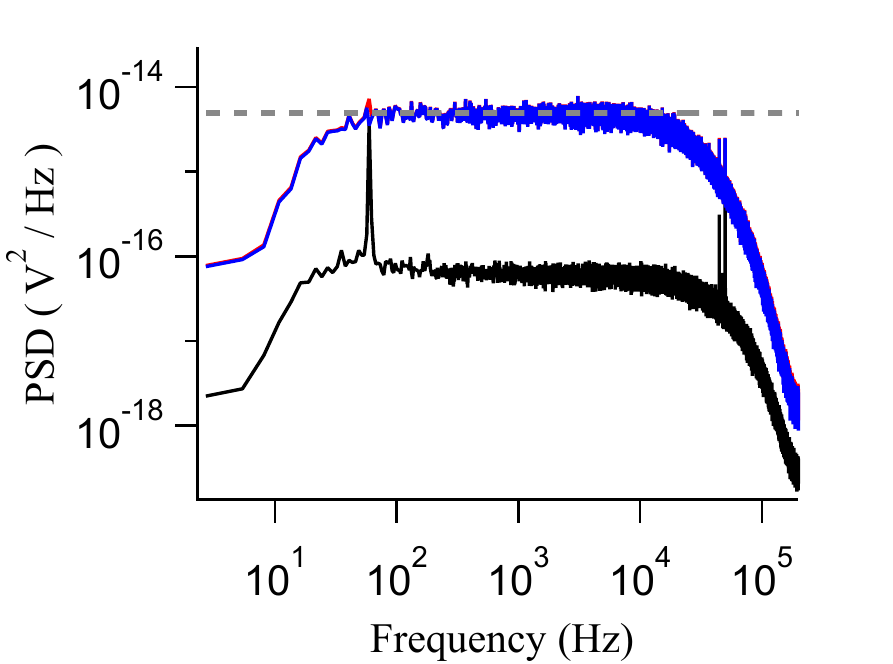}
    \caption{Amplifier noise characterization. Power spectral density of the electronic noise (black), and 300-k$\Omega$ resistor with (red) and without (blue) the background electronic noise.  Note that the red curve is almost covered by the blue one.  Gray dashed line: expected Johnson noise, for $T=298$~K and $R=300$~k$\Omega$.} 
\label{fig:AmpNoise}
\end{figure}

In Fig.~\ref{fig:AmpNoise}, we show the power spectral densities of the background electronic noise (black, using a 1~$\Omega$ resistor, whose Johnson noise is negligible relative to the amplifier noise) and 300-k$\Omega$ resistor (red, with the background noise). The high-pass and low-pass filter values were 10 and 100\,000~Hz, respectively. The latter low-pass cutoff frequency (100~kHz) is higher than the cutoff frequency of the nonequilibrium noise (24~kHz) and thus does not play a significant role in determining the overall noise properties.  The signals were measured using all the amplifiers, with nominal amplifications $G_1 = 600$, $G_2 = 10$, and $G_3 = 15$. The sampling frequency was 400~kHz. The measured output signals were referred back to the respective noise sources by dividing the PSD responses by $(G_1 G_2 G_3)^2$. 

The blue curve represents the background-subtracted power spectral density (PSD) of the 300-k$\Omega$ resistor. The gray line represents the expected Johnson noise spectral density ($=4\,k_\mrm{B}TR$) for a 300-k$\Omega$ resistor, calculated using $R = 300$~k$\Omega$ and $T = 298$~K. We find that experimentally measured response (blue) and expected noise (gray line) are in good agreement. The empirical low- and high-frequency cutoffs are defined as the frequencies where the measured power is reduced by a factor of 1/2. We find $18$~Hz and $24$~kHz, respectively.

We measure the signal-to-noise ratio (SNR) using the PSDs, where the signal is the Johnson noise of the 300-k$\Omega$ resistor and the noise is the background electronic noise. We fit the flat regions of the power spectral densities (black and blue curves) to constants and find from the ratio of the fit values that SNR $\approx$ 76.8.

\begin{figure}[ht!]
    \centering
    \includegraphics[width=0.8\linewidth]{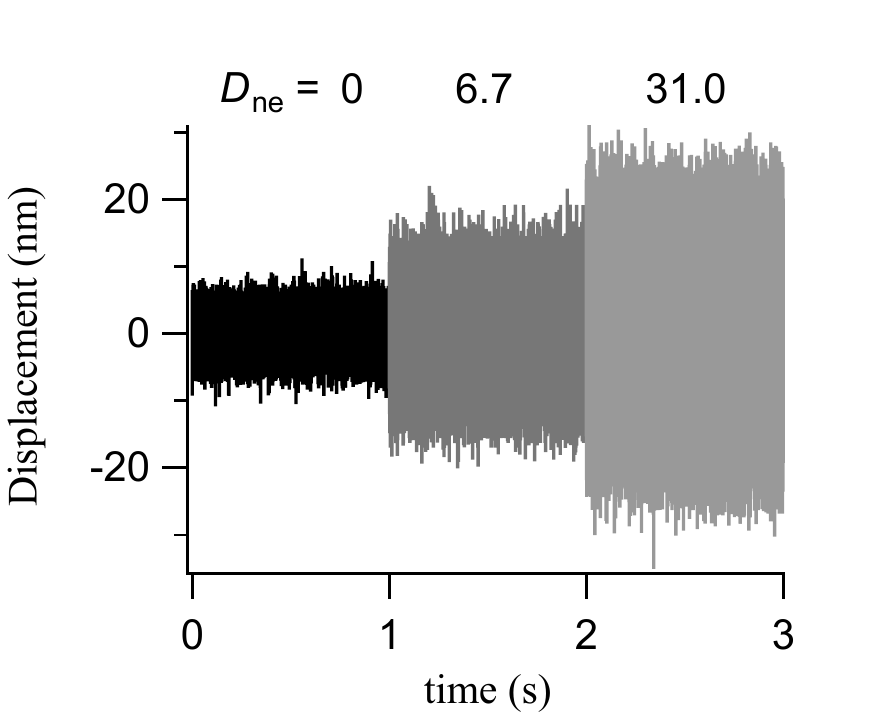}
    \caption{Displacement of the bead ($\xkp-\xk$) in a static trap for $D_\textrm{ne} =$ 0, 6.7, and 31.0.}  
    \label{fig:Fvsfc}
\end{figure}

\section{Numerical simulations}
\label{sec:numerical_sims}
We simulate the dynamics of the engine with nonequilibrium fluctuations using discrete-time dynamics to model the evolution of the bead and the nonequilibrium noise from one time step to the next. To this end, we integrate over one time step the coupled Langevin equations for the bead $x$~[Eq.~(1)]
and for the nonequilibrium noise $\zeta$~[Eq.~(3)].
Then the new particle position $x_{k+1}$ and the next value of the noise $\zeta_{k+1}$ are drawn from a bivariate Gaussian distribution $\mathcal{N}(\mathbf{r};\bm{\mu},\mathbf{\Sigma})$ in $\mathbf{r}$ with mean $\bm\mu$ and covariance matrix $\mathbf\Sigma$. Here,
\begin{widetext}
\begin{align}
    p(x_{k+1},\zeta_{k+1}|x_k,\zeta_{k}) &= \mathcal{N}\left[
    \begin{pmatrix}
    x_{k+1}\\
    \zeta_{k+1}
    \end{pmatrix};\begin{pmatrix}
    \mu_{x_{k+1}}(x_k,\zeta_k)\\ 
    \e^{-\fne \ts} \zeta_k
    \end{pmatrix}, \begin{pmatrix}
    c_{xx} & 
    c_{x\zeta}\\ 
    c_{x\zeta} & c_{\zeta \zeta}
    \end{pmatrix} \right] \label{eq:Gaussian_update}
\end{align}
with mean
\begin{align} 
     \mu_{x_{k+1}}(x_k,\zeta_k) = \left( x_k - \lambda_k + \dg \right) \e^{-\ts} + \lambda_k - \dg
     + \zeta_k \frac{1}{1-\fne} \left( \e^{-\fne \ts} - \e^{-\ts} \right)\,, \label{eq:prop_mean}
\end{align}
and covariances
\begin{subequations}
\begin{align}
    c_{xx} &= \frac{4\Dne\fne^2}{(1-\fne)^2 (1+\fne)}\, \e^{-(1+\fne)\ts}
    - \frac{\Dne\fne}{(1-\fne)^2} \e^{-2\fne\ts}
    -\frac{\Dne\fne^2 + (1-\fne)^2}{(1-\fne)^2} \e^{-2\ts} + \frac{1 + \Dne\fne + \fne}{1+\fne} \label{eq:prop_var} \\
    c_{x\zeta} &= \Dne \Bigg[ \frac{2\fne^2}{1-\fne^2} \e^{-(1+\fne)\ts} - \frac{\fne}{1-\fne} \e^{-2\fne\ts} +\frac{\fne}{1+\fne} \Bigg]\\
    c_{\zeta\zeta} &= \Dne\fne\left( 1-\e^{-2\fne \ts} \right)\,,
\end{align}
\end{subequations}
\end{widetext}

Equation~\eqref{eq:Gaussian_update} was used to simulate the relaxation of the bead in a static potential at $\lambda_k$ and the evolution of the additional Ornstein-Uhlenbeck noise. The trap was updated according to
\begin{align}
    \lambda_{k} = \lambda_{k-1} + 2\,\Theta(x_k - \lambda_{k-1})\,(x_k - \lambda_{k-1})\,, \label{eq:feedback_rule}
\end{align}
where $\Theta(\cdot)$ denotes the Heaviside step function.

To obtain the curve in Fig.~2,
the rate of free-energy gain $\dot F = \Delta F/K\ts = \dg (\lambda_K - \lambda_0)/K\ts$ from $K=10^4$ time steps was averaged over $10^4$ runs. One initial run was used to reach steady-state ratcheting operation.

\section{Nonequilibrium noise strength}

Here we discuss the method used to determine experimentally the strength $\Dne$ of the nonequilibrium noise source. The measurements are done independently from the information-engine experiments. We determine $\Dne$ by measuring the variance of bead displacements $\Delta x := x_{k+1} - x_{k}$ in a static trap in a bath with nonequilibrium-noise cutoff frequency ($\fne = 24~\mrm{kHz})$; 
see Fig.~\ref{fig:AmpNoise}. At that frequency the bead essentially experiences white-noise fluctuations, such that (in scaled units) $\mathrm{Var}(\Delta x) \approx (1+\Dne) \left( 1-\e^{-\ts} \right)$, which in principle permits determination of $\Dne$.

To also account for small corrections arising from the finite frequency of the nonequilibrium noise, we calculate the full expression for the variance of bead displacements. The dynamics of the bead in a static trap with nonequilibrium noise are given by the propagator in Eq.~\eqref{eq:Gaussian_update} and setting $\dg=\lambda_k=0$. Specifically,
\begin{align}
    p(x_{k+1}|x_k,\zeta_k) &= \mathcal{N}\left[x_{k+1};\mu_{x_{k+1}}(x_k,\zeta_k),c_{xx}\right]\,,
\end{align}
a Gaussian with mean $\mu_{x_{k+1}}(x_k,\zeta_k)$ and variance $c_{xx}$ given by Eqs.~\eqref{eq:prop_mean} and \eqref{eq:prop_var}, respectively.

From Eq.~\eqref{eq:Gaussian_update} we also obtain the stationary probability distribution by taking the limit $\ts \to \infty$,
\begin{align}
    \pi(x_k,z_k) = \mathcal{N}\left[\begin{pmatrix}
    x_k\\\zeta_k
    \end{pmatrix};\begin{pmatrix}
    0\\0
    \end{pmatrix},\begin{pmatrix}
    1 + \frac{\Dne\fne}{1+\fne} & \frac{\Dne\fne}{1+\fne}\\
    \frac{\Dne\fne}{1+\fne} & \Dne\fne
    \end{pmatrix}
    \right]\,. \label{eq:stat_dist_trap}
\end{align}

The distribution of the increment given $x_k$ and $\zeta_k$ reads
\begin{align}
    p(\Delta x | x_k,\zeta_k) = \mathcal{N}\left[\Delta x;\mu_x(x_k,\zeta_k)-x_k,c_{xx}\right]\,. \label{eq:cond_prob_incr}
\end{align}
Averaging this distribution over the stationary distribution of $x_k$ and $\zeta_k$ in Eq.~\eqref{eq:stat_dist_trap} yields the average variance of the increment,
\begin{subequations}
\begin{align}
    \mathrm{Var}(\Delta x) &= (e^{-\ts}-1)^2 \left( 1 + \frac{\Dne\fne}{1+\fne} \right)\nonumber\\
    &\quad+  2\frac{(e^{-\ts}-1)}{1-\fne} \left( \e^{-\fne \ts} - \e^{-\ts} \right) \frac{\Dne\fne}{1+\fne}\nonumber\\
    &\quad+ \left[\frac{1}{1-\fne} \left( \e^{-\fne \ts} - \e^{-\ts} \right) \right]^2 \Dne\fne + c_{xx}\\
    &= \frac{2\fne\Dne}{\fne^2 - 1} \left( \fne - 1 - \fne e^{-\ts} + e^{-\fne \ts} \right)\nonumber\\
    &\qquad+ 2\left(1-e^{-\ts}\right)\,.
\end{align}
\end{subequations}

The increase of bead displacement variance relative to the equilibrium case thus reads
\begin{align}
    \frac{\mathrm{Var}(\Delta x)}{\mathrm{Var}_0(\Delta x)} &= 1 + \Dne \frac{\fne \left( \fne - 1 - \fne e^{-\ts} + e^{-\fne \ts} \right)}{\left(\fne^2-1\right) \left(1-e^{-\ts} \right)}\,. 
\label{eq:rel_var_incr}
\end{align}

Equation~\eqref{eq:rel_var_incr} was used to estimate $\Dne$ in experiments, where the typical parameter values were $\ts = 20$~\textmu s $/\tau_\mathrm{r} = 0.025$ and $\fne = 24~\mrm{kHz} / f_\mathrm{c} = 118$.
Note that, in the white-noise limit where $\fne \gg 1$, the right-hand size of Eq.~\eqref{eq:rel_var_incr} reduces to $1+\Dne$, indicating that the position increments result from white-noise fluctuations with a higher effective temperature, in agreement with Eq.~(7) in the main text.

\section{Information efficiency}

Reference~\cite{Ehrich2022_Energetic} shows that to carry out feedback control, a work $W_k^\mathrm{add}$ is needed in addition to the trap work. With no nonequilibrium noise ($\Dne = 0$) and for feedback rules that are independent of the current controller state, i.e., rules of the form $p_\mathrm{c}(z_{k}|x_k)$, with $z_k$ the state of the feedback controller, the additional work is no less than the change in entropy achieved by the control operation,
\begin{subequations}
\label{eq:min_add_control_cost}
\begin{align}
    W_k^\mathrm{add} &\geq k_\mathrm{B} T \left(H[Z_{k-1}|X_k] - H[Z_{k}|X_k] \right) \label{eq:min_add_control_cost_conditional}\\
    &= k_\mathrm{B} T \left(H[X_k,Z_{k-1}] - H[X_k,Z_{k}] \right)\\
    &=: -k_\mathrm{B} T \Delta H^Z_k\,,
\end{align}
\end{subequations}
where $T$ is the temperature of the feedback-controlled system and the controller, and $\Delta H^Z_k$ is the change in joint entropy~\cite[Chap. 2.2]{Cover2006_Elements} due to the dynamics of the controller $Z$. 

Our information engine differs from the description in Ref.~\cite{Ehrich2022_Energetic} in two ways: First, the particle (but not the controller) is subjected to additional colored nonequilibrium noise whose exponential correlation leads to long-time correlations between particle positions, thus violating the Markov assumption. Second, the feedback rule~\eqref{eq:feedback_rule} is \emph{recursive}: the next trap position $\lambda_{k}$ depends not only on the current particle position $x_k$ but also on the current trap position $\lambda_{k-1}$.

The first difference is addressed by focusing on the white-noise limit of the nonequilibrium noise, $\fne \to \infty$, thus ensuring Markovian particle dynamics at an effective  temperature $T_\mathrm{ne}$ which is higher than the temperature $T$ of the controller. The second difference does not impact the minimum additional work in Eqs.~\eqref{eq:min_add_control_cost}, as we show below.

\subsection{Recursive controller architecture}
Reference~\cite{Ehrich2022_Energetic} shows that Eq.~\eqref{eq:min_add_control_cost} is saturated by a control protocol that quasistatically and reversibly carries the conditional controller state from the distribution before the controller update, $p(z_{k-1}|x_k)$, to the distribution after the controller update, $p(z_k|x_k)$, by choosing a time-dependent control potential for the controller state $Z$ such that the conditional controller distributions correspond to the equilibrium distribution under the control potential at the beginning and end of the controller update. The update can be made arbitrarily short yet still reversible when the controller mobility during the update is large.

Here we require recursive controller updates $p_\mathrm{c}(z_k|x_k,z_{k-1})$. The protocol of Ref.~\cite{Ehrich2022_Energetic} cannot be applied directly because, during its update to $p_\mathrm{c}(z_k|x_k)$, all memory of the previous controller state $z_{k-1}$ was lost; however, this can be remedied by adding a temporary auxiliary variable $u$ to the controller, as follows: Starting out in an uncorrelated reference state, the controller state $z_{k-1}$ is copied to the auxiliary variable $u$. Next, the controller update is carried out as described in Ref.~\cite{Ehrich2022_Energetic} with $u$ taking the role of the previous controller state $z_{k-1}$: $p_\mathrm{c}(z_k|x_k,z_{k-1}=u)$. Finally, the auxiliary variable is reset to the initial uncorrelated reference state. Reference~\cite{Wolpert2019_Space-time} argues that such auxiliary, or hidden, variables are generally necessary when implementing computations with Markovian dynamics.

As in Ref.~\cite{Ehrich2022_Energetic}, all steps are carried out quasistatically and reversibly, such that the total heat flow equals the total change in joint entropy,
\begin{subequations}
\begin{align}
    \Delta H^Z_k &:= H[X_k,Z_k] + H_\mathrm{ref}[U] -\left( H[X_k,Z_{k-1}] + H_\mathrm{ref}[U] \right) \\
    &= H[X_k,Z_k] - H[X_k,Z_{k-1}] \,,
\end{align}
\end{subequations}
where $H_\mathrm{ref}[U]$ is the entropy of the auxiliary variable's reference state. Consequently, the additional work needs to at least compensate the decrease in joint entropy by the controller. 

For our information ratchet, a minimal model of the controller state $z_k$ is simply the trap position $\lambda_k$. Using rescaled units and dividing by the sampling time $\ts$ gives the lower bound for the \emph{additional power} required to update the trap position,
\begin{subequations}
\begin{align}
    P_\mathrm{info} &:= - \frac{H[X_k,\Lambda_k] - H[X_k,\Lambda_{k-1}]}{\ts} \\[3pt]
    &= \frac{H[\Lambda_{k-1}|X_k] - H[\Lambda_k|X_k]}{\ts}\,, \label{eq:info_power_2}
\end{align}
\end{subequations}
which we call \emph{information power}, since it is defined purely in terms of the information gathered through the operation of feedback control. It represents the minimum additional power that must be supplied to the engine to make the feedback step thermodynamically consistent. The real experimental equipment dissipates orders of magnitude more power than this lower bound; however, this need not deter us from studying the fundamental limits of this information engine.

\subsection{Steady-state ratcheting}
To evaluate Eq.~\eqref{eq:info_power_2}, we require the conditional distributions $p(\lambda_{k-1}|x_k)$ before the trap update and $p(\lambda_k|x_k)$ after the trap update. 

We use the fact that, after many feedback steps, the distributions of the \emph{relative coordinates}
\begin{subequations}
\begin{align}
    r_k &:= \lambda_k - x_k\\
    r_k^- &:= \lambda_{k-1} - x_k
\end{align}
\end{subequations}
become stationary, i.e.,
\begin{subequations}
\begin{align}
    p(\lambda_k|x_k) &= p(r_k + x_k|x_k) \to \pi(r_k)\\
    p(\lambda_{k-1}|x_k) &= p(r_k^- + x_k|x_k) \to \pi^-(r_k^-)\,.
\end{align}
\end{subequations}
This allows us to rewrite the conditional entropy
\begin{subequations}
\begin{align}
    H[\Lambda_k|X_k] &= -\int \mathrm dx_k\int \mathrm d\lambda_k\,p(x_k,\lambda_k) \ln p(\lambda_k|x_k)\\
    &= -\int \mathrm d\lambda_k p(\lambda_k)\int \mathrm dr_k\,\pi(r_k) \ln \pi(r_k) \label{eq:entropy_rel_dist_same}\\
    &= -\int \mathrm dr_k\,\pi(r_k) \ln \pi(r_k)\\
    &=: H[R_k]\,,
\end{align}
\end{subequations}
in terms of the entropy of the relative variable, $H[R_k]$, and similarly for
\begin{subequations}
\begin{align}
    H[\Lambda_{k-1}|X_k] &= -\int \mathrm dr_k\,\pi^-(r_k^-) \ln \pi^-(r_k^-) \label{eq:entropy_rel_dist_diff}\\
    &=: H[R_k^-] \ .
\end{align}
\end{subequations}

\subsubsection{Approximations for steady-state relative distributions}
It remains to determine the stationary distributions $\pi(r)$ and $\pi^-(r^-)$ for which, unfortunately, no closed-form solutions exist. We can, however, approximate them to low order in the sampling time $\ts$.

With feedback sufficiently fast, every fluctuation over the threshold is immediately caught and the trap moved in response such that the relative coordinate is reflected to the other side of the threshold. Hence, the dynamics relative to the trapping potential can be approximated by diffusion with a reflecting boundary at $r=0$, which has the stationary distribution 
\begin{align}
    \pi(r) &= \begin{cases}
    A\,\exp \left[- \frac{(r-\dg)^2}{2(1+\Dne)}\right] & r \geq 0\\
    0 & r < 0
    \end{cases}\,, \label{eq:Ansatz_stat_dist}
\end{align}
i.e., the equilibrium distribution at relative temperature $1+\Dne$ but truncated at $r=0$. Normalization implies
\begin{align}
    A= \sqrt{\frac{2}{\pi ( 1+\Dne)}} \left[ 1+ \erf \left(\frac{\dg}{\sqrt{2(1+\Dne)}} \right) \right]^{-1}\,. \label{eq:normalization}
\end{align}
This approximation holds to leading order in the sampling time $\ts$, as we now demonstrate.

The dynamics in Eq.~(1) of the main text
in the white-noise ($\fne\to\infty$) limit lead to the propagator
\begin{align}
    p_x(x_{k+1}|x_{k},\lambda_{k}) = \mathcal{N}\Big[x_{k+1}; \mu_x(x_k,\lambda_k;\ts),c(\ts) \Big]\,,
\end{align}
a Gaussian with mean 
\begin{align}
    \mu_x(x_k,\lambda_k;\ts) = (x_{k}-\lambda_{k}) \e^{-\ts} + \lambda_k - (1-\e^{-\ts}) \dg \,
\end{align}
and variance 
\begin{align}
    c(\ts) = (1+\Dne)(1-\e^{-2\ts}) \,. 
\end{align}
In relative coordinates,
\begin{align}
    p_r(r^-|r) &= \mathcal{N}\Big[ r^-;r \e^{-\ts} + (1-\e^{-\ts})\dg, c(\ts) \big]\,.
\end{align}
Thus, the stationary distribution before feedback is
\begin{subequations}
\begin{align}
    \pi^-(r^-) &= \int \mathrm dr\,p_r(r^-|r)\,\pi(r) \\[3pt]
    &= \frac{A}{2}\exp \left[- \frac{(r^- -\dg)^2}{2(1+\Dne)}\right] \label{eq:stat_dist_diff_times} \\
    &\quad\times \left\{ 1 - \erf\left[ \frac{(\dg -r^-)\e^{-\ts} - \dg}{\sqrt{2(1+\Dne)(1-\e^{-2\ts})}} \right] \right\}\,. \nonumber
\end{align}
\end{subequations}

In relative coordinates the feedback rule~\eqref{eq:feedback_rule} is
\begin{align}
    r = r^- - 2 r^- \Theta(r^-)\,.
\end{align}
This amounts to reflecting the negative tail of the distribution $\pi^-(r^-)$ about $r^- = 0$, giving, for $r^- >0$ and to leading order in $\ts$,
\begin{subequations}
\begin{align}
    \pi^-(r) + \pi^-(-r) &= \frac{A}{2}\exp \left[- \frac{(r^- -\dg)^2}{2(1+\Dne)}\right]\nonumber\\
    &\quad\times \Bigg[ 2- \erf\left(\frac{r}{\sqrt{2(1+\Dne)\ts}} \right)\nonumber\\
    &\quad+ \erf\left( \frac{r}{\sqrt{2(1+\Dne)\ts}} \right) \Bigg]\\
    &= \pi(r)\,.
\end{align}
\end{subequations}
Thus, reflecting the stationary distribution recovers the truncated Gaussian in Eq.~\eqref{eq:Ansatz_stat_dist}.

With these approximations, we recover the correct rate of free-energy gain for $\ts \to 0$. Using the fact that $\lambda_k = \lambda_{k-1} + 2(x_k - \lambda_{k-1})$ if $x_k - \lambda_k > 0$, we get
\begin{subequations}
\begin{align}
    \dot F &= \frac{\dg}{\ts} \int_{-\infty}^{x_k} \mathrm d \lambda_{k-1} \, 2(x_k -\lambda_{k-1}) p(\lambda_{k-1}|x_k) \\
    &= -\frac{2\dg}{\ts} \int_{-\infty}^0 \mathrm d r^- \, r^- \,\pi^-(r^-)\,.
\end{align}
\end{subequations}
For small $\ts$, the left tail of $\pi^-(r^-)$ is dominated by the contribution from the error function in Eq.~\eqref{eq:stat_dist_diff_times}, giving
\begin{align}
    \pi^-(r^-) &\approx \frac{A}{2}\exp \left[- \frac{\dg^2}{2(1+\Dne)}\right] \\
    &\quad\times \left\{ 1 - \erf\left[ \frac{(\dg -r^-)\e^{-\ts} - \dg}{\sqrt{2(1+\Dne)(1-\e^{-2\ts})}} \right] \right\} \nonumber
\end{align}
and thus 
\begin{align}
    \dot F &\overset{\ts \to 0}{\longrightarrow} (1+\Dne)\,A\, \exp\left[ -\frac{\dg^2}{2(1+\Dne)} \right]\,.
\end{align}
Together with $A$ from Eq.~\eqref{eq:normalization}, this is Eq.~(9)
of the main text, which itself originates from Ref.~\cite{saha2021maximizing}, where it was calculated for $\Dne=0$ using a different method based on mean first-passage times.

We use $\pi(r)$ [Eq.~\eqref{eq:Ansatz_stat_dist}] to numerically evaluate $H[R_k]$ [Eq.~\eqref{eq:entropy_rel_dist_same}] and $\pi^-(r^-)$ [Eq.~\eqref{eq:stat_dist_diff_times}], then numerically calculate $H[R_k^-]$ [Eq.~\eqref{eq:entropy_rel_dist_diff}], and finally evaluate $P_\mathrm{info}$ in Eq.~\eqref{eq:info_power_2}.

We compare the information power calculated from approximations of the stationary distribution with that calculated from sampled distributions using numerical simulations according to Sec.~\ref{sec:numerical_sims}. Using $10^4$ feedback steps to reach the steady-state regime, we estimated the distributions $\pi(r_k)$ and $\pi(r_k^-)$ from $K=10^6$ feedback steps using histograms with $n=10^2$ bins of uniform width. The entropy was then computed as
\begin{align}
    H[R_k] = -\sum_{i=1}^n \frac{n_i}{K} \ln\left(\frac{n_i}{K\, \mathrm d r}\right)\,,
\end{align}
where $n_i$ is the number of samples in the $i$th bin. We similarly estimated $H[R_k^-]$. (This histogram method of estimating entropy is systematically biased~\cite[App.\ A.8]{Bialek2012_Biophysics}; however, the bias is insignificant due to the large number of samples and vanishes upon taking the difference of two similarly estimated entropies.)

\begin{figure}[htb]
    \centering
    \includegraphics[width=1\linewidth]{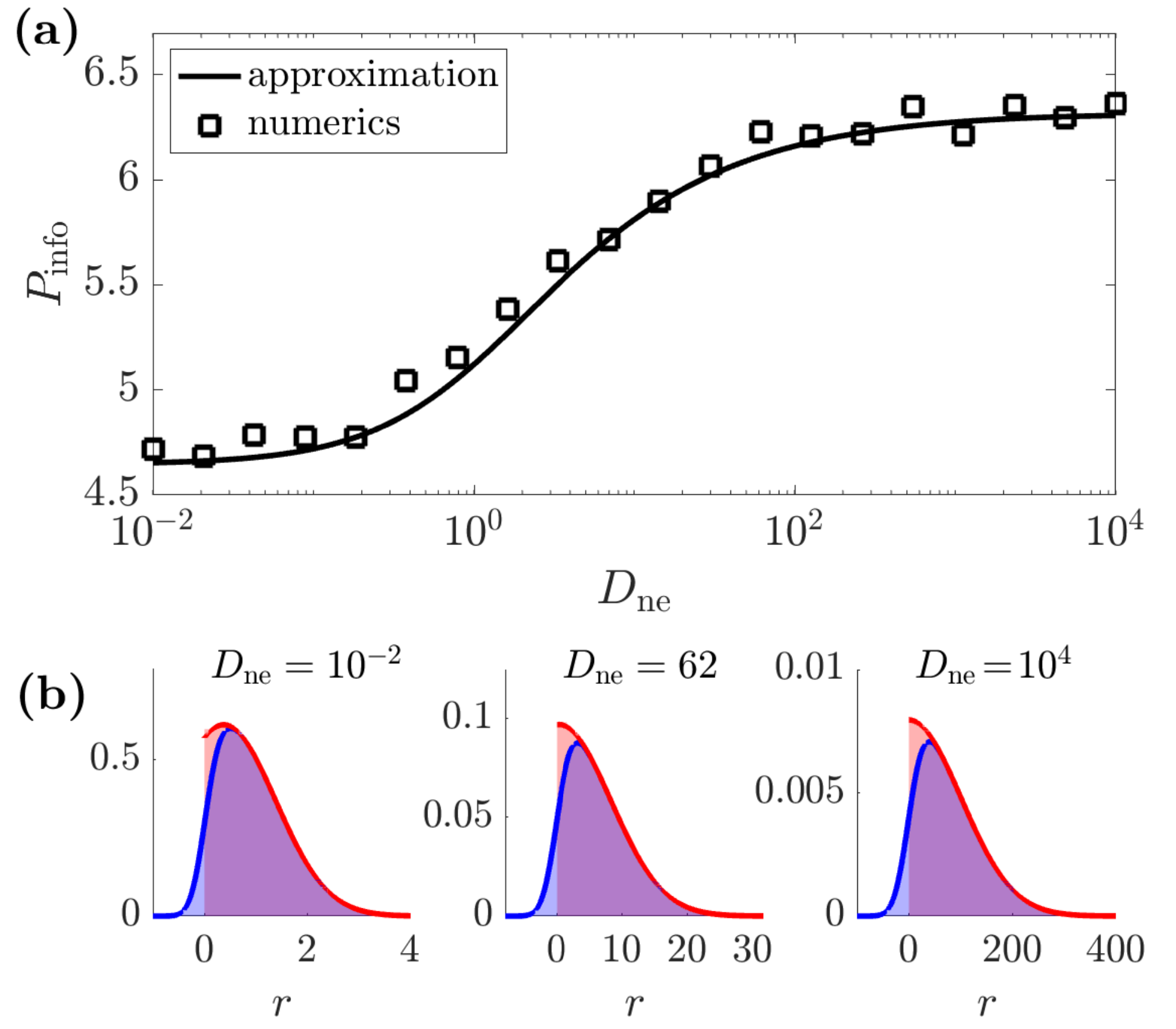}
    \caption{Information power $P_\mathrm{info}$ and distributions of relative coordinate. (a) Information power~\eqref{eq:info_power_2} as a function of nonequilibrium-noise strength $\Dne$. Curve: numerical evaluation of Eqs.~\eqref{eq:entropy_rel_dist_same} and \eqref{eq:entropy_rel_dist_diff} using the analytic approximations in Eqs.~\eqref{eq:Ansatz_stat_dist} and \eqref{eq:stat_dist_diff_times}. Symbols: numerical simulation (Sec.~\ref{sec:numerical_sims}). (b) Stationary relative distributions $\pi(r)$ after feedback (red) and $\pi^-(r^-)$ before feedback (blue), for different nonequilibrium-noise strengths $\Dne$. Curves: analytic approximations [Eqs.~\eqref{eq:Ansatz_stat_dist}~and~\eqref{eq:stat_dist_diff_times}]. Shaded areas: numerical simulations. Parameters are $\ts=1/40$, $\dg =0.38$, and $\fne = 10^4$.}
    \label{fig:information_power}
\end{figure}

Figure~\ref{fig:information_power}(a) shows the information power as a function of the nonequilibrium-noise strength $\Dne$. The approximation matches the estimate from simulations.

Interestingly, the information power saturates in the limit of strong nonequilibrium noise, $\Dne \to \infty$: In this limit, the stationary distributions have a generic shape with a width scaled by the nonequilibrium-noise strength $\Dne$ [see Fig.~\ref{fig:information_power}(b)]. To see this explicitly, consider the limit $\Dne \gg 1$ in Eqs.~\eqref{eq:Ansatz_stat_dist} and \eqref{eq:stat_dist_diff_times},
\begin{subequations}
\begin{align}
    \pi(r) &\to \begin{cases}
    \sqrt{\frac{2}{\pi \Dne}}\,\exp \left[- \frac{1}{2} \left(\frac{r}{\sqrt{\Dne}} \right)^2\right], & r \geq 0\\
    0, & r < 0
    \end{cases}\\
    &= \pi(r/\sqrt{\Dne}) \\
    \pi^-(r^-)
    &= \frac{A}{2}\exp \left[- \frac{(r^-)^2}{2\Dne}\right]\\
    &\quad\times \left\{ 1 + \erf\left[ \frac{r^-\e^{-\ts}}{\sqrt{2\Dne(1-\e^{-2\ts})}} \right] \right\}\nonumber\\
    &= \pi^-(r^-/\sqrt{\Dne})\,.
\end{align}
\end{subequations}
Scaling the width of a distribution by a factor $a$, while keeping it normalized, results in an entropy increase of $\ln |a|$~\cite[Chap. 8.6]{Cover2006_Elements}. Since this scaling affects both distributions $\pi(r)$ and $\pi^-(r^-)$, the entropy difference in Eq.~\eqref{eq:info_power_2} remains unchanged, resulting in saturating information power for $\Dne \to \infty$.

Figure~4(a) in the main text
compares the input information power $P_\mathrm{info}$ with the output free-energy gain $\dot F$ for different nonequilibrium-noise strengths $\Dne$. When the noise is purely thermal, the costs of running the information engine exceed the benefit gained from extracting thermal fluctuations, in accordance with the second law. However, when the nonequilibrium environment supplies sufficiently large fluctuations, the output can dwarf the input by orders of magnitude.

\subsection{Comparison to raising the particle without feedback}
We contrast the information power necessary to run the information engine with the trap power required to drag the particle upwards without feedback. The equation of motion of the particle position $x$ in a trap that is moved upwards at velocity $v$ is
\begin{align}
    \dot x &= -\left[x-\lambda(t)\right] - \dg + \xi(t) +  \zeta(t)\,, \label{eq:Langevin_dragging}
\end{align}
where $\lambda(t) = v t$ is the time-dependent trap position, and $\xi(t)$ and $\zeta(t)$ are the zero-mean thermal white noise and zero-mean nonequilibrium colored noise, respectively (see main text).

Then, the rate of work done on the particle (the trap power) is 
\begin{subequations}
\begin{align}
    \dot W &= \left\langle \partial_\lambda \left\{\frac{1}{2}\left[x-\lambda(t)\right]^2 \right\} \dot\lambda(t) \right\rangle\\
    &= -v \left\langle x- \lambda(t) \right\rangle\\
    &= v^2 + v\,\dg\,, \label{eq:trap_power_dragging_end}
\end{align}
\end{subequations}
where we used the average of Eq.~\eqref{eq:Langevin_dragging} in line~\eqref{eq:trap_power_dragging_end}, $v = \langle \dot x \rangle = - \langle [x-\lambda(t)] \rangle - \dg$.

With $\dot F = v \dg$, we compare the efficiency of the information engine with that of a feedback-free pulling strategy that achieves the same output power:
\begin{subequations}
\begin{align}
    \dot W &= \dot F + \frac{\dot F^2}{\dg^2}\\
    &> \dot F\,,
\end{align}
\end{subequations}
which indicates that the efficiency is bounded by one, $\dot F/\dot W< 1$. Figure~4(b)
in the main text compares the efficiencies of both driving strategies. In contrast to the feedback-free pulling strategy, the information engine can have efficiency (or coefficient of performance) far greater than one.

\section{Simulation code and data}
The code used to perform the simulations and to generate the numerical data can be found in Ref.~\cite{code}.


\providecommand{\noopsort}[1]{}\providecommand{\singleletter}[1]{#1}%

\end{document}